\documentclass[twocolumn,amsmath,amssymb,prd,preprint]{revtex4}

\font\tenrm=cmr10

\def\al{\alpha}
\def\be{\beta}

\def\de{\delta}

\def\et{\eta}

\def\la{\lambda}

\def\rh{\rho}

\def\si{\sigma}

\def\cE{{\cal E}}
\def\cl{{\cal L}}

\def\half{{\textstyle{1\over 2}}}

\def\frac#1#2{{\textstyle{{#1}\over {#2}}}}

\def\lsim{\mathrel{\rlap{\lower4pt\hbox{\hskip1pt$\sim$}}
    \raise1pt\hbox{$<$}}}
\def\gsim{\mathrel{\rlap{\lower4pt\hbox{\hskip1pt$\sim$}}
    \raise1pt\hbox{$>$}}}
\def\sqr#1#2{{\vcenter{\vbox{\hrule height.#2pt
         \hbox{\vrule width.#2pt height#1pt \kern#1pt
         \vrule width.#2pt}
         \hrule height.#2pt}}}}

\def\tr#1{{\rm{tr}}(#1)}

\def\prt{\partial}

\def\pt#1{\phantom{#1}}

\newcommand{\beq}{\begin{equation}}
\newcommand{\eeq}{\end{equation}}
\newcommand{\bea}{\begin{eqnarray}}
\newcommand{\eea}{\end{eqnarray}}
\newcommand{\bit}{\begin{itemize}}
\newcommand{\eit}{\end{itemize}}
\newcommand{\rf}[1]{(\ref{#1})}

\begin{document}

\thispagestyle{empty}

\onecolumngrid

\begin{flushright}
IUHET 482\\
March 2005\\
\end {flushright}

\vskip 1 truein

\begin{center}

{\large\bf Gravity from Local Lorentz Violation}{*}
\\

\bigskip\bigskip\bigskip

V.\ Alan Kosteleck\'y$^{a}$
and Robertus Potting$^b$
\\

\medskip

\it
$^a$Physics Department, Indiana University, Bloomington, IN 47405 
\\
$^b$CENTRA, Physics Department, FCT, 
Universidade do Algarve, 8000 Faro, Portugal
\\

\vskip 1 truein

\rm
{Abstract}
\\

\end{center}

\onecolumngrid

{\rightskip=3pc\leftskip=3pc\noindent

In general relativity,
gravitational waves propagate at the speed of light,
and so gravitons are massless.
The masslessness can be traced to symmetry under diffeomorphisms.
However,
another elegant possibility exists:
masslessness can instead arise 
from spontaneous violation of local Lorentz invariance.
We construct the corresponding theory of gravity.
It reproduces the Einstein-Hilbert action
of general relativity at low energies and temperatures.
Detectable signals occur for sensitive experiments,
and potentially profound implications emerge 
for our theoretical understanding of gravity. 
}

\vskip 1 truein

\begin{flushleft}
\qquad *{\tenrm Third Award,
Gravity Research Foundation, 2005. }
\end{flushleft}

\vfill\eject
\setcounter{page}{1}

Gravity is curvature of spacetime,
and gravitational waves are ripples in its fabric
that transport energy and momentum density.
The ripples propagate at the speed of light,
so their quanta, the gravitons, are massless.
Remarkably,
these features of the ripples contain enough information 
to reconstruct the full theory of general relativity. 
Indeed,
one derivation of the Einstein equations 
is to start with the free theory 
for a massless field $h_{\mu\nu}$ 
propagating in Minkowski spacetime,
and then to impose a self-consistent coupling 
to the energy-momentum tensor $T_{\mu\nu}$.
The self-consistency requirement
leads uniquely to the Einstein equations,
and the resulting series of corrections to the Minkowski metric
sums to give the Riemann metric and the familiar spacetime geometry
\cite{expgr,sd}.

In this derivation,
the reason for starting with 
a symmetric field $h_{\mu\nu}$ 
is readily understood,
since it is needed in the action to couple 
to the symmetric energy-momentum tensor $T_{\mu\nu}$.
But why are the gravitons massless?

Masslessness is often taken to be the consequence
of a symmetry.
In quantum electrodynamics the masslessness of the photon 
is normally attributed to gauge invariance,
or symmetry under local changes of phase.
In quantum chromodynamics,
the theory of the strong interaction,
masslessness of the gluons is likewise attributed
to a gauge invariance,
albeit a nonlinear one.
In general relativity,
the masslessness of gravitons
can be traced to symmetry under active diffeomorphisms:
no diffeomorphism-invariant mass term exists.

Nature and mathematics allow,
however, 
for an alternative reason why a field might be massless.
Surprisingly,
this alternative explanation involves the breaking
of a symmetry rather than its existence.
A general result,
the Nambu-Goldstone theorem
\cite{ng},
states under mild assumptions that 
there must be a massless particle
whenever a continuous global symmetry of an action 
isn't a symmetry of the vacuum.
This result is readily understood for the simple case
of an action with a kinetic term $K$ 
and a nonderivative potential $V$.
The vacuum solution has zero $K$ and is a minimum of $V$.
By assumption, 
a symmetry of the theory transforms any given minimum of $V$
to a different minimum, 
so $V$ has at least one flat direction.
Small vibrations perpendicular 
to the flat direction are massive modes,
with mass related to the curvature of the potential at the minimum,
but vibrations parallel to it are massless because
the potential is flat.
The corresponding massless particles are called Nambu-Goldstone modes.

In this essay,
we show that an alternative description of gravity 
can be constructed from a symmetric two-tensor 
without the assumption of masslessness.
In this picture,
masslessness is a consequence of symmetry breaking 
rather than of exact symmetry:
diffeomorphism symmetry and local Lorentz symmetry 
are spontaneously broken,
but the graviton remains massless because it is a Nambu-Goldstone mode.
Remarkably,
the Einstein-Hilbert action of general relativity is recovered
at low energies and temperatures.
However,
the new theory is the same as general relativity 
only in this limit,
so differences between the two theories
can emerge under suitable circumstances.
It follows that the origin of the graviton's masslessness 
is an issue with potentially profound theoretical implications
and is experimentally testable.

To derive the new theory,
we adopt the construction
beginning with a basic action in Minkowski spacetime 
and then imposing self-consistency 
of the coupling to the energy-momentum tensor.
The cardinal object in the theory is a symmetric two-tensor,
denoted by $C_{\mu\nu}$.
The Lagrange density for the basic theory is 
\bea
\cl_C &=&
\half 
C^{\mu\nu} K_{\mu\nu\al\be} C^{\al\be} - V(C_{\mu\nu}).
\label{clag}
\eea
Here,
$K_{\mu\nu\al\be}$ is
the usual quadratic kinetic operator for a massless spin-2 field,
and the potential $V$ is a scalar functional of $C_{\mu\nu}$.
This theory is invariant 
under local Lorentz transformations
and under diffeomorphisms.

The vacuum is found by minimizing $V$.
The functional dependence of $V$ on $C_{\mu\nu}$ 
is chosen to ensure that the cardinal field 
has a constant nonzero vacuum value
$C_{\mu\nu} = c_{\mu\nu}$,
where
$c^{\mu\nu}c_{\mu\nu} = c^2$
is a real number.
This spontaneously breaks 
local Lorentz and diffeomorphism invariances.
The massless Nambu-Goldstone fields are the excitations 
$\de C_{\mu\nu} \equiv C_{\mu\nu} - c_{\mu\nu}$
about this solution,
generated by the broken symmetries
and maintaining the potential minimum.
In what follows,
we show these excitations play the role 
of the graviton field $h_{\mu\nu}$.

How many massless fields are generated in this way?
Since there are six Lorentz transformations 
(three rotations and three boosts)
and four diffeomorphisms,
there could in principle be up to 10 massless modes:
six Lorentz modes contained 
in an antisymmetric field $\cE_{\mu\nu}$,
and four diffeomorphism modes contained
in a field $\Xi_\mu$.
However,
the kinetic operator in the Lagrange density 
is purely quadratric in derivatives,
a feature known to imply that the diffeomorphism field
$\Xi_\mu$ decouples from the action 
at leading order in small fields
\cite{bk}.
This can be confirmed directly:
denoting by $D_\mu$
the covariant derivative in general coordinates 
in Minkowski spacetime,
at leading order in small fields we find
$D_\la C_{\mu\nu} \approx
\prt_\la \cE^{\pt{al}\al}_{\mu} c_{\al\nu}
+ \prt_\la \cE^{\pt{al}\al}_{\nu} c_{\al\mu}$,
which depends only on the Lorentz field $\cE_{\mu\nu}$.
We can therefore write 
\beq
\de C_{\mu\nu} 
\equiv M_{\mu\nu}^{\pt{\mu\nu}\al\be} \cE_{\al\be} 
= \cE^{\pt{al}\al}_{\mu} c_{\al\nu} 
+ \cE^{\pt{al}\al}_{\nu} c_{\al\mu} ,
\label{eps}
\eeq
where
$M_{\mu\nu\al\be} = 
\half 
( \et_{\mu\al} c_{\nu\be} 
+ \et_{\nu\al} c_{\mu\be} 
- \et_{\mu\be} c_{\nu\al}
- \et_{\nu\be} c_{\mu\al} 
)$.
Since there are only six independent fields in $\cE_{\mu\nu}$,
the 10 independent components of $\de C_{\mu\nu}$
must satisfy four identities.
For generic $c_{\mu\nu}$,
these  are 
\beq
\tr{\de C~c^m} = 0,
\label{decid}
\eeq
with $m = 0,1,2,3$.

The properties of the candidate graviton field $\de C_{\mu\nu}$ 
are given by its equations of motion.
Varying the Lagrange density with respect to 
the independent degrees of freedom $\cE_{\mu\nu}$ gives 
$ M^{\mu\nu\rh\si} K_{\mu\nu\al\be} \de C^{\al\be} =0$.
These equations can be solved using Fourier decomposition.
The solutions obey
the usual massless wave equation, 
\beq
\prt^\la\prt_\la \de C_{\mu\nu} = 0,
\label{redepeqmot}
\eeq
subject to the Hilbert condition
$\prt^\mu \de C_{\mu\nu}=0$.
The latter imposes another four constraints 
on $\de C_{\mu\nu}$.
This means only two combinations of 
the massless Lorentz modes $\cE_{\mu\nu}$ propagate.

These results are promising:
they show that for weak fields
the new theory contains the expected number of modes 
obeying the massless wave equation.
However,
if the new theory indeed contains general relativity,
the weak-field limit of the new theory 
must match exactly the usual description 
of a massless graviton field $h_{\mu\nu}$ in Minkowski spacetime.
How does this match arise?

To answer this,
let's start with the Lagrange density 
for a free massless graviton field $h_{\mu\nu}$,
\beq
\cl_h =
\half 
h^{\mu\nu} K_{\mu\nu\al\be} h^{\al\be} .
\label{hlag}
\eeq
Initially,
$h_{\mu\nu}$ has 10 degrees of freedom.
Since the theory is diffeomorphism invariant
and there is no potential $V$ to drive spontaneous symmetry breaking,
diffeomorphisms can be used to make four coordinate choices.
These choices represent a gauge restriction on 
$h_{\mu\nu}$.
Most studies of gravitational waves
adopt the so-called transverse-traceless gauge,
but here we pick instead a different gauge 
that yields a direct match to the new theory.
For generic $c_{\mu\nu}$,
we choose the four conditions
\beq
\tr{h~c^m} = 0
\label{newg}
\eeq
with $m = 0,1,2,3$.
A short calculation verifies that this is an acceptable gauge choice.

The behavior of the graviton field $h_{\mu\nu}$
is fixed by the equations of motion,
$K_{\mu\nu\al\be} h^{\al\be} = 0$.
These can be solved by 
performing a Fourier decomposition 
and using the specific gauge choice.
The solutions 
obey the Hilbert condition $\prt^\mu h_{\mu\nu}=0$
and satisfy the wave equation 
\beq
\prt^\la \prt_\la h_{\mu\nu} = 0.
\label{redheqmot}
\eeq
They describe two graviton degrees of freedom 
propagating as massless waves,
as usual.

We can now see how the match between the two theories arises.
Both start with symmetric fields,
$h_{\mu\nu}$ and $\de C_{\mu\nu}$.
In the usual picture,
diffeomorphism symmetry permits some components 
of $h_{\mu\nu}$ to be fixed,
while in the new theory
only six components of $\de C_{\mu\nu}$ are independent
from the start.
In both cases,
the Hilbert condition holds 
and the two graviton degrees of freedom
obey the massless wave equation.
At this level,
the equations for the two theories
are in direct correspondence,
even though the symmetry structures of the two theories
are radically different.

How can the full nonlinear Einstein equations be obtained?
In the usual $h_{\mu\nu}$ theory,
this can be done by adding a coupling 
to the energy-momentum tensor $T_{\mu\nu}$
and imposing self-consistency of its conservation.
An elegant single-step procedure has been given by Deser
\cite{sd}.
For our purposes,
the key feature of this derivation 
is its gauge independence,
which means our specific gauge choice \rf{newg} for $h_{\mu\nu}$ 
still allows construction of the Einstein-Hilbert action.
Since the two theories match 
when this gauge is used,  
it follows that the new theory
contains general relativity too, 
via the same construction.

We have thus come to the surprising result
that the conventional description of gravity 
via general relativity
follows as a consequence of spontaneous breaking 
of local Lorentz symmetry.
This result may seem paradoxical,
since the existence of local Lorentz symmetry 
is among the underlying foundations of general relativity.
However,
the paradox is superficial:
although the new theory contains general relativity,
the two theories are different.

Do the differences lead to physical effects?
These are still being explored,
but the answer is definitely yes.
The new theory includes various corrections 
to the action of general relativity.
Effects arise in both the pure-gravity sector 
and the matter sector,
and also from the structural differences.
Let's take a brief tour 
of some results obtained so far.
 
In the pure-gravity sector,
the new theory has subleading corrections 
to the Einstein equations in vacuum,
which generate potentially observable effects
in sensitive experiments.
For nearly Minkowski spacetimes
and hence in laboratory or solar-system tests,
including gravitational-wave experiments,
the effects are nonzero but small because 
they enter at higher order in small fields.
However,
in more extreme contexts 
such as black holes or the early Universe,
the deviations from general relativity can be significant.

Effects in the matter sector arise because the coupling 
of the cardinal field to the energy-momentum tensor
generates an extra term $c_{\mu\nu}T^{\mu\nu}$ in the action. 
This has potential physical implications 
for the behavior of matter in laboratory searches 
for Lorentz breaking,
including ones that aren't traditionally viewed
as gravitational tests. 
For example,
terms of this type can modify signals 
in clock-comparison experiments,
in neutrino oscillations,
and in tests with light,
among others. 
A framework for the comprehensive treatment 
of such effects exists
\cite{akgrav},
and numerous searches are presently under way.

Structural differences between the two theories
include the nonzero vacuum value $c_{\mu\nu}$,
which may help explain dark energy,
and effects at energies and temperatures near the Planck scale.
At high energies,
the extra degrees of freedom in the cardinal field 
become relevant because the modes perpendicular to the flat direction
of the potential $V$ can be excited.
High temperatures restore the local Lorentz symmetry
because $V$ acquires corrections 
\cite{dj}
that change its shape,
making stable a zero value for $C_{\mu\nu}$.
The former gravitons then become oscillations about this new minimum
and acquire large masses.
This may have profound implications
for the very early Universe.
 
The quantum properties of gravity are also significantly changed,
which may shed light on the perennial
problem of the consistent quantization of gravity.
For example,
recent results for vector fields 
\cite{bak}
can be adapted for the new theory 
to show that nonpolynomial 
and superficially unrenormalizable potentials $V$
can become renormalizable and stable 
when quantum corrections are included.
These theories have spontaneous Lorentz violation,
so the requirements of renormalizability and stability 
of the basic theory \rf{clag}
suffice to ensure a massless graviton 
without diffeomorphism symmetry,
a surprising result.

In our derivation leading to the new description of gravity,
the geometrical properties of the theory are obscured,
just as the geometry is obscured
in the similar construction of general relativity.
A geometrical formulation should exist
and is actively being sought.
It must be both beautiful and unconventional,
describing gravity as spacetime curvature
and tying spacetime curvature to rippling cardinal fields
coupled in a self-consistent way.

This work was supported in part
by the Department of Energy
under grant number DE-FG02-91ER40661,
by the National Aeronautics and Space Administration
under grant number NAG3-2914,
and by the 
Funda\c{c}\~ao para a Ci\^encia e a Tecnologia (Portugal).

\end{document}